\documentclass[12pt,amsfonts]{article}
\begin{document}
\def\p {{\partial}}
\def\n {{\nu}}
\def\m {{\mu}}
\def\a {{\alpha}}
\def\bt {{\beta}}
\def\f {{\phi}}
\def\th {{\theta}}
\def\g {{\gamma}}
\def\eps {{\epsilon}}
\def\e {{\psi}}
\def\la {{\lambda}}
\def\na {{\nabla}}
\def\k {\chi}
\def\bn {\begin{eqnarray}}
\def\en {\end{eqnarray}}
\title{Lagrangian formulation of classical fields within Riemann-Liouville fractional derivatives}
\maketitle
\begin{center}
\author{
\textbf{Dumitru Baleanu} \vspace{0.2cm}\footnote{On leave of
absence from Institute of Space Sciences, P.O.BOX, MG-23, R 76900,
Magurele-Bucharest, Romania,E-mails: dumitru@cankaya.edu.tr,
baleanu@venus.nipne.ro}\\
Department of Mathematics and Computer Sciences, Faculty of Arts
and Sciences, \c{C}ankaya University- 06530, Ankara, Turkey}\\
\vspace{0.5cm}

 {\textbf{Sami I. Muslih}} \vspace{0.2cm}\footnote{E-mail:
smuslih@ictp.trieste.it}\\Department
of Physics, Al-Azhar University, Gaza, Palestine\\and\\
International Center for Theoretical Physics(ICTP),\\Trieste,
Italy
\end{center}
\hskip 5 cm
\begin{abstract}
 The classical fields with fractional derivatives are investigated by using the
 fractional Lagrangian formulation.The fractional Euler-Lagrange equations were obtained and  two examples
 were studied.
\end{abstract}

\newpage

\section{Introduction}

Fractional derivatives \cite{samko,olham,podlubny}, have played a
significant role in physics, mathematics and engineering during
the last decade \cite{hilfer,machado,klimek,agrawal, agrawal3}.
Fractional calculus found many interesting applications in recent
studies of scaling phenomena \cite{nonnenmacher, metzler,
zaslavsky} or in classical mechanics
\cite{riewe1,riewe2,rabei,agrawal1,agrawal2}.

 Riewe  has used the fractional calculus to develop a formalism which can be
used for both conservative and non conservative systems
\cite{riewe1,riewe2}. Although many laws of nature can be obtained
using certain functionals and the theory of calculus of
variations, not all laws can be obtained by using this way. For
example, almost all systems contain internal damping, yet
traditional energy based approach cannot be used to obtain
equations describing the behavior of a non conservative system
\cite{riewe1,riewe2}. Using the fractional approach, one can
obtain the Euler-Lagrange and the Hamiltonian equations of motion
for the nonconservative systems. The simple solutions of the
fractional Dirac equation of order $\frac{2}{3}$ were investigated
recently \cite{raspini1,raspini2}.Even more recently the
fractional variational principle in macroscopic picture was
discussed in \cite{agaies}.

 Recently, an extension of the simplest fractional problem and the
fractional variational problem of Lagrange was obtained
\cite{agrawal1,agrawal2}. A natural generalization of Agrawal's
approach, was to apply the fractional calculus to constrained
systems \cite{dirac} and to obtain both the fractional
Euler-Lagrange equations \cite{baleanu1} and the fractional
Hamiltonian formulation of constrained systems \cite{baleanu2,
muslih}.

The fractional Lagrangian is non-local, therefore we should take
care of this property in handling with its corresponding
Hamiltonian. An interesting proposal for the Hamiltonian formalism
corresponding to the non-local Lagrangian systems was considered
in \cite{llosa}. The physical degrees of freedom of non-local
theories was investigated very recently in \cite{gomis}.
Besides,the Hamiltonian formalism for non-local field theories
\cite{seiberg} in d space-time dimensions was developed recently
in \cite{gomis1}.

 For these reasons the fractional variational problems for fields are interesting
to be investigated.

 The aim of this
paper is to obtain the Euler-Lagrange equations for the classical
fields with Riemann-Liouville fractional derivatives.

The present paper is organized as follows. In section 2 the
fractional Euler-Lagrange equations for fields are obtained. In
section 3 the fractional Klein-Gordon equations and Dirac's
equation of order $\frac{2}{3}$ are obtained.Finally, the section
4 is dedicated to our conclusions.

\section{ Lagrangian  formulation of field systems with
fractional derivatives}

\subsection{Riemann-Liouville partial fractional derivative}

Let us consider a function f depending on n variables,
$x_1,x_2,\cdots x_n$.A partial left Riemann-Liouville fractional
derivative of order $\alpha_k$, $0<\alpha_k<1$, in the k-th
variable is defined as [1,2] \begin{equation} ({\textbf{D}_{a_k
+}^{\alpha_k}}f)(x)=\frac{1}{\Gamma{(1-\alpha_k)}}\frac{\partial}{\partial
x_k}\int_{a_k}^{x_k}
\frac{f(x_1,\cdots,x_{k-1},u,x_{k+1},\cdots,x_n)}{(x_k-u)^{\alpha_k}}du\end{equation}

and a partial right Riemann-Liouville fractional derivative of
order $\alpha_k$ has the form

\begin{equation} ({\textbf{D}_{a_k
-}^{\alpha_k}}f)(x)=-\frac{1}{\Gamma{(1-\alpha_k)}}\frac{\partial}{\partial
x_k}\int_{x_k}^{a_k}
\frac{f(x_1,\cdots,x_{k-1},u,x_{k+1},\cdots,x_n)}{(-x_k+u)^{\alpha_k}}du\end{equation}

If the function f is differentiable we obtain

\begin{eqnarray}\label{ecue3}
({\textbf{D}_{a_k
+}^{\alpha_k}}f)(x)&=&\frac{1}{\Gamma(1-\alpha_k)}[\frac{f(x_1,\cdots,x_{k-1},a_k,x_{k+1},\cdots,x_n)}{(x_k-a_k)^{\alpha_k}}\cr
&+& \int_{a_k}^{x_k}\frac{\frac{\partial f}{\partial
u}(x_1,\cdots,x_{k-1},u,x_{k+1},\cdots,x_n)}{(x_k-u)^{\alpha_k}}]du.
\end{eqnarray}

We notice that the last term of the equation (\ref{ecue3})
represents the Caputo derivative \cite{caputo}. This derivative
has the advantage that certain initial conditions are easier to
interpret.

\subsection{Fractional classical fields}

A covariant form of the action would involve a Lagrangian density
${\cal L}$ via $S=\int {\cal L}d^{4}x= \int {\cal L} d^{3}x dt$
where ${\cal L} = {\cal L}(\phi,\partial_\mu\phi)$ and with $L=
\int {\cal L} d^3x$. The corresponding covariant Euler-Lagrange
equations are

\begin{equation}
\frac{\partial {\cal L}}{\partial\phi} -\partial_\mu\frac{\partial
{\cal L}}{\partial(\partial_\mu\phi)}=0,
\end{equation}

where $\phi$ is the field variable .

 Now we shall investigate the
fractional generalization of the above Lagrangian density.

Let us consider the action function of the form

\begin{equation}
S=\int {\cal L}\left(\phi(x),({\textbf{D}_{a_k -}^{\alpha_k}})
\phi(x),({\textbf{D}_{a_k +}^{\alpha_k}})\phi(x), x\right)d^3xdt.
\end{equation}
Here $0 < \a_k\leq 1$ and $a_k$ correspond to $x_1,x_2,x_3$ and t
respectively. For each partial derivative we may have a specific
order $\alpha_k$ and a given limit $a_k$.

In this paper the integration limits are  $-\infty$ and $\infty$
respectively.Under this conditions ${\textbf{D}_{a_k
+}^{\alpha_k}}$ would become ${\textbf{D}_{-\infty +}^{\alpha_k}}$
and ${\textbf{D}_{a_k -}^{\alpha_k}}$ would become
${\textbf{D}_{\infty -}^{\alpha_k}}$.

Let us consider the $\epsilon$ finite variation of the functional
$S(\phi)$, that we write with explicit dependence from the fields
and their fractional derivatives

\begin{eqnarray}\label{ecu}
 \Delta_{\epsilon} S(\phi)&=&\int
 [{{\cal L}(x,\phi+\epsilon\delta\phi, ({\textbf{D}_{\infty -}^{\alpha_k}})
\phi(x)+\epsilon({\textbf{D}_{\infty
-}^{\alpha_k}}})\delta\phi,({\textbf{D}_{-\infty +}^{\alpha_k}})
\phi(x)\cr &+&\epsilon({\textbf{D}_{-\infty
+}^{\alpha_k}})\delta\phi) - {\cal L}(x,\phi,({\textbf{D}_{\infty
-}^{\alpha_k}}) \phi(x),({\textbf{D}_{-\infty +}^{\alpha_k}})
\phi(x))]d^3xdt.
\end{eqnarray}

We will develop the first term in square brackets, which is a
function on $\epsilon$ as a Taylor series in $\epsilon$, stopping
at the first order. Therefore from (\ref{ecu}) we obtain

\begin{eqnarray}\label{actiune}
\Delta_{\epsilon} S(\phi) &=& \int [{\cal L}(x,\phi,
({\textbf{D}_{\infty -}^{\alpha_k}}) \phi(x),({\textbf{D}_{\infty
+}^{\alpha_k}}) \phi(x))+(\frac{\partial {\cal
L}}{\partial\phi}\delta\phi)\epsilon\cr
 &+&\sum_k^{1,4} \frac{\partial
{\cal L}}{\partial({\textbf{D}_{\infty
-}^{\alpha_k}}\phi)}\delta({\textbf{D}_{\infty
-}^{\alpha_k}}\phi)\epsilon + \sum_k^{1,4} \frac{\partial {\cal
L}}{\partial({\textbf{D}_{-\infty
+}^{\alpha_k}}\phi)}\delta({\textbf{D}_{-\infty
+}^{\alpha_k}}\phi)\epsilon +O(\epsilon)\cr &-&{\cal L}(x,\phi,
({\textbf{D}_{\infty -}^{\alpha_k}})\phi(x),({\textbf{D}_{-\infty
+}^{\alpha_k}})\phi(x))]d^3xdt
\end{eqnarray}

Taking into account (\ref{actiune}) the form of (\ref{ecu})
becomes

\begin{eqnarray}\label{delta1}
 \Delta_{\epsilon} S(\phi)&=&\epsilon\int [ \frac{\partial
{\cal L}}{\partial\phi}\delta\phi+\sum_k^{1,4} \frac{\partial
{\cal L}}{\partial({\textbf{D}_{\infty
-}^{\alpha_k}}\phi)}({\textbf{D}_{\infty
-}^{\alpha_k}}\delta\phi)\cr &+& \sum_k^{1,4} \frac{\partial {\cal
L}}{\partial({\textbf{D}_{-\infty
+}^{\alpha_k}}\phi)}({\textbf{D}_{-\infty
+}^{\alpha_k}}\delta\phi) +O(\epsilon) ]d^3xdt.
\end{eqnarray}

We now perform a fractional integration by parts of the second
term in (\ref{delta1}) by using the formula \cite{samko, love}

\begin{equation}
\int_{-\infty}^{\infty}f(x)({\textbf{D}_{-\infty
+}^{\alpha_k}}g)(x)
dx=\int_{-\infty}^{\infty}g(x)({\textbf{D}_{\infty
-}^{\alpha_k}}f)(x)dx.
\end{equation}

Therefore we obtain

\begin{eqnarray}
 \Delta_{\epsilon} S(\phi)&=&\epsilon\int [ \frac{\partial
{\cal L}}{\partial\phi}\delta\phi+\sum_k^{1,4}
\{({\textbf{D}_{-\infty +}^{\alpha_k}})\frac{\partial {\cal
L}}{\partial({\textbf{D}_{\infty
-}^{\alpha_k}})\phi}\}\delta\phi\cr &+& \sum_k^{1,4}
\{({\textbf{D}_{\infty -}^{\alpha_k}})\frac{\partial {\cal
L}}{\partial({\textbf{D}_{-\infty +}^{\alpha_k}})\phi}\}\delta\phi
]d^3xdt + \int O(\epsilon)d^3xdt.
\end{eqnarray}

After taking the limit $lim_{\epsilon\longrightarrow
0}\frac{\Delta_{\epsilon} S(\phi)}{\epsilon}$ we obtain the
fractional Euler-Lagrange equations as follows

\begin{equation}\label{fel}
\frac{\partial {\cal L}}{\partial\phi}+\sum_k^{1,4}
\{({\textbf{D}_{-\infty +}^{\alpha_k}})\frac{\partial {\cal
L}}{\partial({\textbf{D}_{\infty -}^{\alpha_k}})\phi}+
({\textbf{D}_{\infty -}^{\alpha_k}})\frac{\partial {\cal
L}}{\partial({\textbf{D}_{-\infty +}^{\alpha_k}})\phi}\}=0.
\end{equation}

 It is worth
commenting that for $\a_k \rightarrow 1$, the equations
(\ref{fel}) are the usual Euler-Lagrange equations for classical
fields.

\section{Examples}

In this section we shall analyze two systems. The first one is the
Klein-Gordon  field and the second one is the Dirac field.

\subsection{Fractional Klein-Gordon equation}

 As a first example let us consider the following Lagrangian
density as

\begin{equation}\label{kg}
{\cal L_{KG}}=\frac{1}{2}[\partial_\mu\phi\partial^{\mu}\phi
-m^2\phi^2]
\end{equation}

or

\begin{equation}\label{kg1}
{\cal L_{KG}}=\frac{1}{2}[(\dot\phi)^2- (\nabla\phi)^2
-m^2\phi^2].
\end{equation}

The fractional generalization of (\ref{kg1}) is given by

\begin{equation}
{\cal L_{FKG}}=\frac{1}{2}[({\textbf{D}_{-\infty
+}^{\alpha_t}\phi})^2-({\textbf{D}_{-\infty
+}^{\alpha_x}\phi})^2-({\textbf{D}_{-\infty
+}^{\alpha_y}\phi})^2-({\textbf{D}_{-\infty +}^{\alpha_z}\phi})^2
-m^2\phi^2].
\end{equation}

By using (\ref{fel}), the fractional Euler-Lagrange equations are
obtain as follows

\begin{equation}
 \textbf{D}_{\infty-}^{\alpha_t} (\textbf{D}_{ -\infty +}^{\alpha_t})\phi-
\textbf{D}_{ \infty -}^{\alpha_x}(\textbf{D}_{ -\infty
+}^{\alpha_x})\phi-\textbf{D}_{\infty -}^{\alpha_y}(\textbf{D}_{
-\infty+}^{\alpha_y})\phi-\textbf{D}_{
\infty-}^{\alpha_z}(\textbf{D}_{ -\infty +}^{\alpha_z})\phi-
m^2\phi=0.
\end{equation}

For all $\a_k\rightarrow 1$ we obtain the usual Klein-Gordon
equation.

\subsection{Fractional Dirac equation}

 The Dirac Lagrangian is given by

\begin{equation}
{\cal L_D}={\bar\psi}(i\gamma^{\mu}\partial_\mu-m)\psi.
\end{equation}

 In \cite{raspini1,raspini2} a detailed analysis of the solutions of
 the fractional Dirac equation of $\frac{2}{3}$ was done.

We propose the Lagrangian density for Dirac field of order
$\frac{2}{3}$ as follows

\begin{equation}\label{dd}
{\cal L_{FD}} = {\bar\psi}\left(\gamma^{\a} \textbf{D}_{\a-}^{2/3}
\psi(x) + (m)^{2/3}\right)\psi(x).
\end{equation}

Taking into account (\ref{fel}) and (\ref{dd}) the corresponding
Dirac equation becomes

\begin{equation}\label{ddd}
\gamma^{\a}\textbf{D}_{\a+}^{2/3}\Psi(x) + (m)^{2/3}\Psi(x)=0.
\end{equation}

We would like to stress that in (\ref{dd}) and (\ref{ddd}) the
expressions of $\textbf{D}_{\a-}^{2/3}$ and
$\textbf{D}_{\a+}^{2/3}$ have the same meaning as it was
introduced in the previous paragraph but for the sake of
simplicity we kept the compact form for summations $\gamma^{\a}
\textbf{D}_{\a-}^{2/3}$ and $\gamma^{\a}\textbf{D}_{\a+}^{2/3}$
respectively.

This result is the same as the fractional Dirac equation obtained
in references \cite{raspini1,raspini2}.

\section{Conclusion}

The fractional Lagrangian is not unique due to the fact that we
have several choices to replace the fractional derivatives in the
usual one.This property of the fractional Lagrangian is an
advantage of this theory. For a specific problem we have several
options to obtain, under the limit process, the usual
Euler-Lagrange equations for fields.

In this paper we have extended the derivation of the usual
Euler-Lagrange equations of motion for classical field to the case
the Lagrangian contains fractional derivatives of fields. This
method has been applied with the variational principle to obtain
the corresponding fractional Euler-Lagrange equations. The
fractional Klein-Gordon equations were obtained by using the
fractional  variational principle. This approach allows  to obtain
the Dirac equation with fractional derivatives of order $2/3$
recently obtained in \cite{raspini1,raspini2}.

\section*{Acknowledgments}

D.B would like to thank J.A.Machado for many stimulating
discussions. This work is partially supported by the Scientific
and Technical Research Council of Turkey.

\end{document}